\documentclass[aps,prb,twocolumn,superscriptaddress,showpacs,showkeys]{revtex4}
\usepackage{amsmath,amssymb}
\usepackage{graphicx}
\usepackage{subfigure}
\def\RR{\mathbb{R}} 

\def\PP{\mathbb{P}}
\def\<{\langle} \def\>{\rangle}

\newcommand{\uu}[1]{{\boldsymbol #1}}
\def\vv{\uu{v}}

\def\xx{\uu{x}}

\def\vv{\uu{v}}
\def\zz{\uu{z}}

\def\eeta{\uu{\eta}}

\usepackage{amsthm}
\begin{document}

\title{Exact rate calculations by trajectory parallelization and
  twisting} \author{Eric Vanden-Eijnden} \email{eve2@cims.nyu.edu}
\affiliation{%
  Courant Institute of Mathematical Sciences, New York University, New
  York, NY 10012, USA} 
\author{Maddalena Venturoli}
\email{mventuro@cims.nyu.edu} 
\affiliation{%
  Courant Institute of Mathematical Sciences, New York University, New
  York, NY 10012, USA}           

\date{\today}

\begin{abstract} A sampling procedure to compute exactly the rate of
  activated processes arising in systems at equilibrium or
  nonequilibrium steady state is presented.
  The procedure is a generalization of the method
  in [A.~Warmflash, P.~Bhimalapuram, and A.~R. Dinner, J.\ Chem.\
  Phys. {\bf 127}, 154112 (2007); A.~Dickson, A.~Warmflash, and A.~R.
  Dinner, J.\ Chem.\ Phys. {\bf 130}, 074104 (2009)] in which one
  performs simulations restricted into cells by using a reinjection
  rule at the boundaries of the cells which is consistent with
  the exact probability fluxes through these boundaries. Our
  generalization uses results from transition path theory which
  indicate how to twist the dynamics to calculate reaction rates.
\end{abstract}


\maketitle

\paragraph*{Introduction.}
The main objective of this work is to revisit and extend in scope the
nonequilibrium sampling procedure to compute steady state probability
distributions proposed in
Refs.~\onlinecite{warmflashJCP07,dicksonJCP09}. Specifically, we show
how this procedure can be modified to calculate exactly certain
dynamical quantities such as the rate of reactions occurring in
arbitrary equilibrium or nonequilibrium systems at statistical steady
state. This is done by exploiting results of transition path theory
(TPT)~\cite{tpt2,tpt_erice,tpt_ex} which indicate how to twist the
dynamics of a given system to calculate the reaction rate between a
given reactant and product state. The method proposed in this note can
be seen as a generalization to arbitrary nonequilibrium processes at
statistical steady state of the milestoning procedure with Voronoi
tessellation proposed in Ref.~\onlinecite{miles6} by building upon the
original works in Refs.~\onlinecite{miles1,miles2,miles3}. This
generalization makes the procedure more expensive computationally, but
it permits to relax completely the assumptions made in milestoning --
these assumptions were discussed in detail in
Ref.~\onlinecite{miles5}. Our method can also be viewed as a
generalization of the transition interface sampling
(TIS)~\cite{tis2,pptis,tis3} and forward flux sampling
(FFS)~\cite{ffs,valerianiFFS} methods in which arbitrary sets of
interfaces can be used that do not have to be placed in monotone
succession.

The remainder of this note is organized as follows. First we revisit
from an original perspective the nonequilibrium sampling procedure of
Refs.~\onlinecite{warmflashJCP07,dicksonJCP09}. Next, we show how this
procedure can be modified to calculate reaction rates exactly using
TPT. We then compare our procedure with the Markovian milestoning
method proposed in Ref.~\onlinecite{miles6} and with
TIS~\cite{tis2,pptis,tis3} and FFS~\cite{ffs,valerianiFFS}. Finally we
illustrate our procedure on a simple example. In terms of notations
and assumptions, we will denote by $\zz$ the location of the system in
its state-space $\Omega\subset \RR^d$ (e.g.~it could be the positions
and velocities of all the atoms in a molecular system, in which case
$\zz= (\xx,\vv)$ and $d=6n$ if $n$ is the number of atoms). The
specifics of the dynamics of the system are not important except that
we assume that (i) its evolution is Markovian and (ii) it is ergodic
with respect to a probability density function which we denote
by~$\varrho(\zz)$. Notice that we do not require detailed balance,
i.e.~$\varrho(\zz)$ is associated with a nonequilibrium statistical
steady state in general.

\paragraph*{Restricted sampling with flux matching.}

\begin{figure}[thbp]
   \centerline{\includegraphics[width=3in]{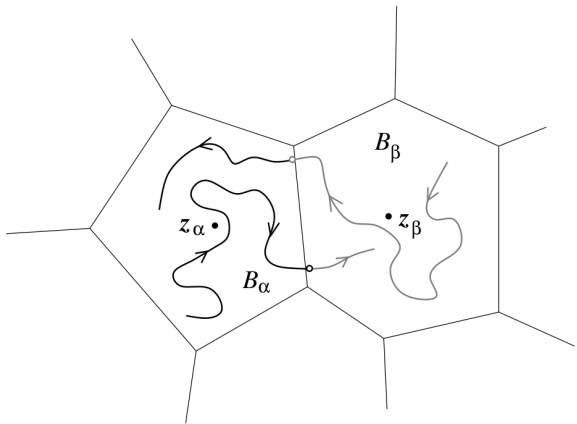}}
   \caption{%
Representation of a portion of the Voronoi tessellation
in two-dimension, where point $\zz_\alpha$ is the generator of cell
$B_\alpha$ and point $\zz_\beta$ the generator of cell $B_\beta$. 
Pieces of restricted trajectory inside $B_\alpha$
are shown in black and inside $B_\beta$ in grey. Each time the
trajectory in one cell attempts to exit the cell, the exit point
position (depicted as an open circle) is stored, and used to
reinitialize the trajectory into the neighboring cell. }
\label{fig:voronoi}
\end{figure}

The methods in Refs.~\onlinecite{warmflashJCP07,dicksonJCP09,miles6}
are based on a factorization of the dynamics in which one artificially
constrains the system to evolve in a set of cells partitioning
state-space in a way that (i) does not bias the dynamics inside the
cells and (ii) is consistent with the exact probability fluxes in and
out of these cells. In other words, the procedure guarantees that a
true unconstrained trajectory of the system can be reconstructed
exactly by patching together in some appropriate way the pieces
computed in the cells. These pieces can be calculated in parallel in
each cell, hence the name trajectory parallelization. The method in
Ref.~\onlinecite{miles6} is restricted to equilibrium systems and
exploits the time-reversibility of the dynamics. The method in
Refs.~\onlinecite{warmflashJCP07,dicksonJCP09} is more costly but it
works also for nonequilibrium systems. To explain how the latter
method works, we first recall how to construct the cells using the
Voronoi tessellation associated with a given set of generating points
or centers~\cite{miles6,fts}. Denoting these centers by $\zz_{\alpha}
\in \Omega \subset \RR^d$, with $\alpha=1,\ldots,\Lambda$, the Voronoi
cell $B_\alpha$ associated to $\zz_{\alpha}$ contains all the points
that are closer to $\zz_\alpha$ than to any other center, i.e.~(see
Fig.~\ref{fig:voronoi} for an illustration)
\begin{equation}
  \label{eq:Bk}
  B_{\alpha}= 
  \{\zz \in \Omega \; : \; 
  \|\zz - \zz_{\alpha}\|<\|\zz-\zz_{\beta}\| \; 
  \text{for all} \; \beta \ne \alpha  \},
\end{equation}
where $\|\cdot\|$ is some appropriate norm (e.g.~the Euclidean norm in
which case $\|\zz\|^2=\sum_{i=1}^d z_i^2$).

If we were to generate an infinitely long trajectory of the system,
$\zz(t)$ with $t>0$, this trajectory would keep going in and out of
the cells $B_\alpha$ by crossing the edges between these cells. Out of
this trajectory we could therefore generate an ensemble of exit-entry
points, i.e.~those points on the edges of the cells at which the
trajectory goes from a cell $B_\alpha$ into a neighboring cell
$B_{\beta}$. By the Markovian assumption, these points are all we need
to generate an exact sample of trajectories inside each of the cells
$B_\alpha$ simply by starting trajectories at the points leading into
that cell and running these trajectories forward in time until they
exit the cell. The procedure in
Refs.~\onlinecite{warmflashJCP07,dicksonJCP09} is a way to generate
these pieces of trajectories inside the cells without having to
compute the entry points beforehand from a long unbiased trajectory
but rather by generating them on-the-fly. To understand how this is
done, imagine that we associate an independent copy, or replica, of
the system to each cell $B_\alpha$. Let us denote the instantaneous
position of these replicas by $\zz_\alpha(t)\in B_\alpha$,
$\alpha=1,\ldots, \Lambda$. Even if we start $\zz_\alpha(t)$ inside
$B_\alpha$, sooner or later this trajectory will try to exit
$B_\alpha$ and go to another cell. When this happens, we store the
exit point on the boundary, put the trajectory on hold and wait until
a trajectory in one of the neighboring cells makes an attempt to exit
this cell by crossing the boundary with $B_\alpha$. We then take this
crossing point and use it to reinitialize the trajectory in
$B_\alpha$. This is illustrated in Fig.~\ref{fig:voronoi}. At the
beginning of the simulation we do not have enough re-entry points, so
many of the replicas may be on hold. But as the simulation goes on, we
can build databanks of re-entry points from any $B_\beta$ into any
$B_\alpha$ (assuming that these two cells have a common boundary --
otherwise the databank is trivially empty), which contain the last $X$
points by which the trajectory tried to escape from cell $B_\beta$ and
enter cell $B_\alpha$. This way, each time a trajectory tries to exit
a cell $B_\alpha$, we can immediately pick a re-entry point into that
cell from the appropriate databank and continue the trajectory from
that point without having to put it on hold.

The only remaining issue we need to take care of in order to make
complete the procedure outlined above is how to pick the re-entry
point into $B_\alpha$ among the databanks on the various edges leading
into $B_\alpha$. By construction, the re-entry points in the databanks
on each edge are unbiased samples of re-entry points conditional on
the trajectory entering by that edge. But there are several edges by
which the trajectory can enter a given cell, and to introduce no bias
we need to pick the edges with the proper probability of re-entrance
by that edge. To see how this can be done, imagine that, as we run the
simulation in each cell $B_\alpha$, we compute an estimate of the
effective rate of exit out that cell and into $B_\beta$ via
\begin{equation}
  \label{eq:fluxdef}
  \nu_{\alpha,\beta} = \frac{N_{\alpha,\beta}}{T_\alpha}
\end{equation}
where $N_{\alpha,\beta}$ is the total number of times the trajectory
hit the boundary between $B_\alpha$ and $B_\beta$ (which is also the
number of times the trajectory in cell $B_\alpha$ had to be re-injected
into that cell from a re-entry point) and $T_\alpha$ is the total
simulation time in cell $B_\alpha$ (i.e.~the total time the trajectory
in cell $B_\alpha$ was running). If we then denote by $\pi_\alpha$ the
probability to find the unbiased trajectory inside cell $B_\alpha$ at
statistical steady state, i.e.
\begin{equation}
  \label{eq:rhocells}
  \pi_{\alpha} = \int_{B_\alpha} \varrho(\zz) d\zz,
\end{equation}
we see that $\pi_\alpha$ and $\nu_{\alpha,\beta}$ are related as
\begin{equation}
  \label{eq:peq}
  \sum_{\substack{\beta=1\\ \beta \not=\alpha}
  }^{\Lambda} \pi_{\beta} \nu_{\beta,\alpha} = 
  \sum_{\substack{\beta=1\\ \beta \not=\alpha}}^{\Lambda} 
  \pi_{\alpha} \nu_{\alpha,\beta},\qquad 
  \sum_{\alpha=1}^{\Lambda} \pi_\alpha = 1. 
\end{equation}
The first equation in~\eqref{eq:peq} simply expresses that, at
statistical steady state, the total probability flux into $B_\alpha$
(which is the term at the left hand-side of the first equation
in~\eqref{eq:peq}) must be equal to the total flux out of $B_\alpha$
(which is the term at the right hand-side of the first equation
in~\eqref{eq:peq}). The second equation in~\eqref{eq:peq} is simply a
normalization condition for the probability which follows from
$\sum_{\alpha} \pi_\alpha = \sum_{\alpha} \int_{B_\alpha} \varrho(\zz)
d\zz = \int_\Omega \varrho(\zz) d\zz = 1$. We stress that
\eqref{eq:peq} does not require detailed balance (i.e.~$\pi_{\beta}
\nu_{\beta,\alpha} \not = \pi_{\alpha} \nu_{\alpha,\beta}$, possibly)
and so it holds even for nonequilibrium processes provided only that
they are at statistical steady state. Having calculated
$\nu_{\alpha,\beta}$ from~\eqref{eq:fluxdef} and $\pi_\alpha$
from~\eqref{eq:peq}, we then have an estimate for the probability flux
from $B_\beta$ into $B_\alpha$: $\pi_{\beta} \nu_{\beta,\alpha}$.
Consistently, the probability that the trajectory enters cell
$B_\alpha$ by coming from $B_\beta$ is simply:
\begin{equation}
  \label{eq:probflux}
  \PP_{\partial B_{\beta} \cap \partial B_{\alpha}}=
  \frac{\pi_{\beta} \nu_{\beta,\alpha}}
  {\sum_{\beta'\neq\alpha} \pi_{\beta'} \nu_{\beta',\alpha} }, 
  \quad (\beta\neq \alpha),
\end{equation}
if $B_\alpha$ and $B_\beta$ have a common edge and $\PP_{\partial
  B_{\beta} \cap \partial B_{\alpha}}=0$ otherwise. This expression
gives us the desired probability to pick the edge~$\partial B_{\beta}
\cap \partial B_{\alpha}$ for re-entry into $B_\alpha$.

Summarizing, the algorithm to perform simulations restricted inside
the cells for systems at nonequilibrium steady state is as follows:

\begin{itemize}
\item[ (1)] Denoting by $\zz_{\alpha}(t)\in B_\alpha$  the current
  state of
  the replica in cell $B_\alpha$, let $\zz_{\alpha}^\star$ be the
  state of the system produced from
  $\zz_{\alpha}(t)$ after one
  timestep~$\mathit{\Delta} t$ by a standard (i.e.~unrestricted)
  integrator for the system
  (e.g.~velocity-Verlet if the system is a molecular dynamics (MD)
  system). If $\zz^\star_{\alpha}\in
  B_{\alpha}$, set
\begin{equation}
  \label{eq:reject}
    \zz_{\alpha}(t+\mathit{\Delta} t) = \zz^\star_{\alpha}.
\end{equation}
Otherwise, if $\zz^\star_{\alpha}\in B_{\beta}$ with $\beta \neq \alpha$:
\begin{itemize}
\item[ (i)] Store the point $\zz_{\alpha}^\star$
  in a databank of entry points from $B_{\alpha}$ into $B_{\beta}$. In
  other words,
  assuming that the databank contains already $m$ points
  $\zz_{\alpha,\beta}^{k}$ with
  $k=1,2,\ldots,m$, set
  $\zz_{\alpha,\beta}^{m+1}=\zz_{\alpha}^\star$. 
\item[ (ii)] Update $\nu_{\alpha,\beta}$ via \eqref{eq:fluxdef},
  $\pi_{\alpha}$ via~\eqref{eq:peq} and $\PP_{\partial B_{\beta} \cap
    \partial B_{\alpha}}$ via~\eqref{eq:probflux}.

\item[(iii)] Select an edge $\partial B_{\beta'} \cap \partial
  B_{\alpha}$ of $B_{\alpha}$ with probability proportional to
  $\PP_{\partial B_{\beta'} \cap
    \partial B_{\alpha}}$. 
\item[(iv)] Pick a point $\zz_{\beta',\alpha}^{k}$ with uniform
  probability from the databank of re-entry points on edge 
  $\partial B_{\beta'} \cap \partial
  B_{\alpha}$, and set
\begin{equation}
  \label{eq:reinject}
  \zz_{\alpha}(t+\mathit{\Delta} t) = \zz_{\beta',\alpha}^{k} 
\end{equation}
\end{itemize}

\item[(2)] Go to step (1) and iterate to collect statistics.
\end{itemize}

In essence, the algorithm above is the same as the one proposed in
Ref.~\onlinecite{warmflashJCP07,dicksonJCP09}, though the two differ
in the details. For instance, we use a single set of cells instead of
two (in Refs.~\onlinecite{warmflashJCP07,dicksonJCP09} two staggered
sets were used to ensure stability, but we observed no such stability
problems with the procedure above). Besides a set of trajectories, one
of the output of the procedure is to give the probability $\pi_\alpha$
to find the system in cell $B_\alpha$ at statistical steady state
(see~\eqref{eq:rhocells}). Indeed the computation of $\pi_{\alpha}$
was the main objective in
Refs.~\onlinecite{warmflashJCP07,dicksonJCP09}. Below we will show how
to extract more from the procedure, in particular reaction rate
information, by appropriate modifications. Before getting there,
however, let us make a few remarks about the algorithm above. In step
(iv) we have assumed that the list of entry points from $B_{\beta'}$
into $B_{\alpha}$ is not empty and, as already mentioned above, this
may not be true at the beginning of the simulation (there might have
been no collision with that edge up to that time). In that case we
have to put the simulations in some of cells on hold at the beginning
until we get proper re-entry points and start building databanks on
the edges. Note that, at any given time, these databanks could each
contain the last re-entry point only, though in that case we may run
out of points again and have to put replicas on hold. Thus it is safer
to always keep as many points in the databanks as memory allows.
However, the databanks do not have to be enlarged indefinitely -- at
worst we trade memory for CPUs since, the smaller the databanks, the
higher the probability that one replica will have to be put on hold
for some time. Also note that the quantities used to evaluate the
probability in Eq.~\eqref{eq:probflux} have to be computed on-the-fly,
and need information from all the cells. This, again, may lead to
problems at the beginning of the simulations, when the statistics for
$N_{\alpha,\beta}$ is not accurate enough. To overcome this problem we
set $\pi_{\alpha}=cst$ $\forall \alpha=1,\ldots,N$ at the beginning
when statistics is insufficient to solve Eq.~\eqref{eq:peq}. Using
more educated guesses is possible too. Also, it is worth noting that
$\nu_{\alpha,\beta}$ and $\pi_{\alpha}$ can be monitored on-the-fly to
assess their convergence as a function of the length of the
simulation, and the actual simulations of the replicas in each cell
can be performed in parallel; only the re-entry events require
communication. Finally, we should stress that the procedure above
relies on the ability to count the successive points at which a
trajectory crosses the edges of the cells. This may lead to
difficulties if the dynamics is governed by a stochastic differential
equation, in which case these crossing points may form a fractal set.
It leads to no difficulty, however, if some components of the
trajectory are smooth and the cells are defined accordingly. We will
come back to this issue later in the illustrative example section.

\begin{figure}[thbp]
   \centerline{\includegraphics[width=3in]{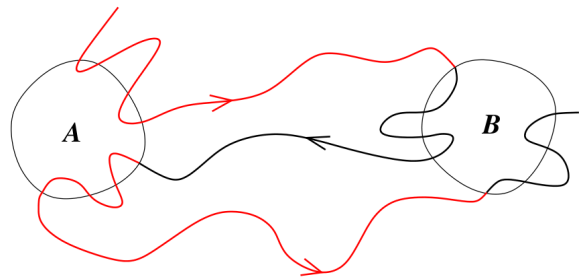}}
   \caption{%
     Schematic representation of a piece of ergodic trajectory
visiting the two sets $A$ and $B$.
     The pieces of this trajectory for which the last visited set was
$A$ are depicted in red, and
     those for which the last visited set was $B$ are depicted in
black.
   }
\label{fig:trajAB}
\end{figure}

\paragraph*{Rate calculation.} Let us now come to the question of how
to compute the reaction rate between a reactant and product state,
which we identify as two disjoint sets in the system's state-space
denoted as $A \subset \Omega$ and $B \subset \Omega$ respectively. As
mentioned earlier, this calculation will be done by twisting the
dynamics in some appropriate way that is dictated by the results of
TPT. We begin by introducing the relevant objects that we will
consider. If we assume again that we have at our disposal an
infinitely long trajectory, $\zz(t)$ with $t>0$, this trajectory will
go back and forth between $A$ and $B$ as time goes on, and we can split
this trajectory into two pieces, depending on whether it visited last
$A$ or $B$. This construction is illustrated in Fig.~\ref{fig:trajAB},
where the trajectory is shown in red if it visited $A$ last and in
black if it visited $B$ last: if the trajectory visited $A$ last at
time $t$, we will say that it is assigned to $A$ at time $t$, if it
visited $B$ last at time $t$, we will say that it is assigned to $B$
at time $t$.

Based on this assignment, we can introduce the following quantities.
If $N^T_{A,B}$ denotes the total number of times the trajectory went
from being assigned to $A$ to being assigned to $B$ during the
time interval $[0,T]$ (i.e.~the number of times it switched from red to
black in Fig.~\ref{fig:trajAB}), we set
\begin{equation}
  \label{eq:nuAB}
  \nu_R=\lim_{T\to\infty}  \frac{N^T_{A,B}}{T}.
\end{equation}
This quantity gives the average frequency at which the trajectory goes
from $A$ to $B$ which, because the system is at steady state, is also
the same as the average frequency at which the trajectory goes from
$B$ to $A$ since $N^T_{A,B} = N^T_{B,A}$ asymptotically. The rate
in~\eqref{eq:nuAB} is referred to as the rate of the reactive
trajectories in TPT, hence the subscript $R$. The rate of reactive
trajectories $\nu_R$ should not be confused with the two reaction
rates from $A$ to $B$ and $B$ to $A$ defined respectively as
\begin{equation}
  \label{eq:kAB} 
  k_{A,B}= \lim_{T\to\infty} \frac{N^T_{A,B}}{T_A}, \quad 
  k_{B,A}= \lim_{T\to\infty} \frac{N^T_{A,B}}{T_B},
\end{equation}
where $T_A$ and $T_B$ are, respectively, the total times during which
the trajectory was assigned to $A$ or $B$ in the interval $[0,T]$
(i.e. the total times the trajectory is red or black in
Fig.~\ref{fig:trajAB}). If $A$ and $B$ are metastable (i.e.~if the
trajectory commits to each of these sets and looses memory of its past
before going back to the other set), $k_{A,B}$ and $k_{B,A}$ are the
rates that enter the phenomenological mass-action law describing how
the populations in $A$ and $B$ evolve in time. Notice that $\nu_R$ is
related to $k_{A,B}$ and $k_{B,A}$ as \begin{equation}
  \label{sec:nuRkABkBA}
  k_{A,B}=\frac{\nu_R }{\rho_A}, \qquad k_{B,A}=\frac{\nu_R}{ \rho_B},
\end{equation}
where $ \rho_A$ is the fraction of time the trajectory is assigned to $A$
and $\rho_B$ the fraction of time it is assigned to $B$:
\begin{equation}
  \label{eq:rhoABrhoBA}
  \rho_{A}=\lim_{T\to \infty} \frac{T_A}{T}, \qquad 
  \rho_{B}=\lim_{T\to \infty} \frac{T_B}{T}.
\end{equation}
By definition $\rho_A\le1$, $\rho_B\le1$ and $\rho_A+\rho_B=1$. The
quantities $\nu_R$, $k_{A,B}$, $k_{B,A}$, $\rho_A$ and $\rho_B$ are
the ones we will show how to compute exactly. To see how this can be
done, it is useful to give first the TPT expressions for these
quantities. These expressions involve the backward committor function,
$q_-(\zz)$, which gives the probability that a trajectory observed at
point $\zz$ is coming from $A$ last rather than from $B$ (i.e.~that it
is assigned to $A$ rather than $B$ using the jargon introduced above).
By definition $q_-(\zz)=1$ if $\zz \in A$, $q_-(\zz)=0$ if $\zz \in
B$, and $0\le q_-(\zz)\le1$ otherwise. The backward committor function
is useful because by the Markovian assumption it follows that, at
statistical steady state, the probability density to observe a
trajectory at point $\zz \in \Omega$ at any given time $t$
\textit{and} that this trajectory is assigned to $A$ at that time is
simply $\varrho_{A}(\zz)= \varrho(\zz) \, q_-(\zz)$. Similarly the
probability density to observe a trajectory at point $\zz \in \Omega$
at time $t$ \textit{and} that this trajectory is assigned to $B$ at
that time is $\varrho_{B}(\zz) = 1-\varrho_A(\zz)= \varrho(\zz) \,
(1-q_-(\zz))$. Since $\rho_A = \int_\Omega \varrho_{A}(\zz)d\zz$ and
$\rho_B = \int_\Omega \varrho_B(\zz)d\zz$ by definition, this
implies that
\begin{equation}
  \label{eq:rhoAB}
  \rho_A=\int_{\Omega} \varrho(\zz) \, q_-(\zz) \, d\zz \le 1, \qquad
  \rho_B=1-\rho_A.
\end{equation}
Similarly, it is easy to see that $\nu_R$ is the total probability
flux associated with $\varrho_{A}(\zz)= \varrho(\zz) \, q_-(\zz)$
going through any dividing surface between $A$ and $B$ (like e.g.~the
boundary of $B$). The explicit form of this flux depends on the
specifics of the dynamics and it is given in Ref.~\onlinecite{tpt_erice}:
let us omit to repeat this formula here since it will not be
important in what follows.

\begin{figure}[thbp]
   \centerline{\includegraphics[width=3in]{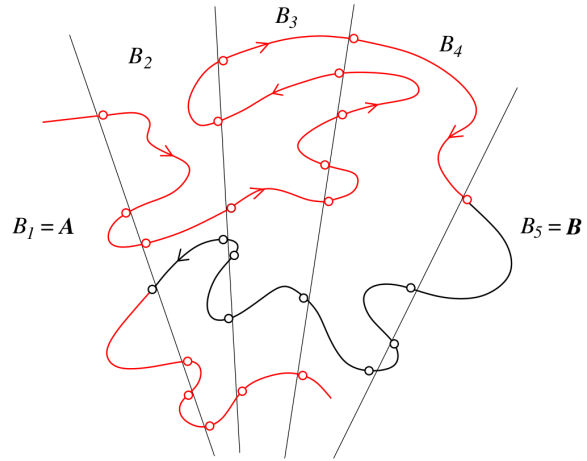}}
   \caption{%
     Schematic illustration of a piece of trajectory
     crossing the cells $B_\alpha$ with $\alpha=1,\ldots,5$,
     where cell $B_1$ is identified with the reactant state $A$ and
     cell $B_5$ with the product state $B$.
     The circles represent the entry points which would be used to
perform sampling restricted in the cells.
     If only the red points are used as re-entry points, then the restricted
     sampling selects the pieces of this trajectory when it is assigned to~$A$ (these pieces are depicted in red).  }
\label{fig:trajABmiles}
\end{figure}
In practice, we do not know explicitly $q_-(\zz)$ (nor even
$\varrho(\zz)$ in most nonequilibrium systems) but we can still make
use of the observations above to modify the sampling procedure
explained before. Suppose that we define the reactant state $A$ as
being the union of a group of cells $B_\alpha$ and the product state
$B$ as the union of another group. Clearly, what we would then like to
do is modify the sampling procedure in such a way that only the
re-entry points associated with the trajectory when it is assigned to
$A$ are put in the databanks (see the illustration in
Fig.~\ref{fig:trajABmiles}) and keep track of the associated
probability fluxes through the boundary of the cells. Indeed, using
these re-entry points and these fluxes only, we would then simulate in
the cells pieces of trajectories that are statistically
indistinguishable from the unbiased trajectory when it is assigned to
$A$. We could then compute the total probability in the cells to get
an estimate of $\rho_A$ (and hence $\rho_B=1-\rho_A$) as well as the
total flux into $B$ to get $\nu_R$. We claim that there is a simple
procedure to do these operations in practice. The key observation is
that trajectories assigned to $A$ can be propagated like regular
trajectories while they are outside of $A$ and $B$ -- the only
constraint imposed on them is via a boundary condition in their past:
they need to have come from~$A$ rather than~$B$ last. But we know what
this boundary condition entails, at least in a statistical sense:
indeed the probability density $\varrho_A(\zz)$ that a trajectory be
at $\zz$ and be assigned to $A$ is equal to the statistical steady
state probability density $\varrho(\zz)$ for $\zz \in A$ while it is
equal to 0 for $\zz \in B$. We also know that the probability flux out
of~$A$ is the statistical steady state one and the probability flux
out of~$B$ is identically zero.

We can easily impose these boundary conditions in practice by
modifying our sampling procedure as follows. Suppose that we have
performed a sampling as before and computed the steady state
(unbiased) $\pi_\alpha$ and $\nu_{\alpha,\beta}$. We can then run
another independent sampling where we only consider the cells outside
of $A$ and $B$. In these cells, we run trajectories as before (though,
as we will see, using re-entry points that are different from the ones
calculated before), store their exit points from the cells to build
databanks of re-entry points in other cells, and compute
\begin{equation}
  \label{eq:fluxdefA}
  \nu^A_{\alpha,\beta} = \frac{N^A_{\alpha,\beta}}{T^A_\alpha}
\end{equation}
where $N^A_{\alpha,\beta}$ is the total number of times the
trajectories hit the boundary between $B_\alpha$ and $B_\beta$ and
$T^A_\alpha$ is the total simulation time in cell $B_\alpha$. Note
that $\nu^A_{\alpha,\beta} $ is only defined by~\eqref{eq:fluxdefA} if
$\alpha$ is the index of a cell $B_\alpha$ not inside $A$ and $B$ (the
index $\beta$, on the other hand, runs over all the cells, including
those forming $A$ and $B$). Consistent with the bias we need to impose
to focus on trajectories assigned to $A$, we supplement this by
$\nu_{\alpha,\beta}^A=\nu_{\alpha,\beta}$ if $\alpha$ is the index of
a cell $B_\alpha$ used to define~$A$ (since the effective rate of exit
out of $A$ must be the unbiased statistical steady one) and by
$\nu_{\alpha,\beta}^A=0$ if $\alpha$ is the index of a cell $B_\alpha$
used to define~$B$ (since the effective rate of exit out of $B$ must
be zero). We also set $\pi^A_\alpha=\pi_\alpha$ if $\alpha$ is the
index of a cell $B_\alpha$ used to define~$A$, $\pi^A_\alpha=0$ if
$\alpha$ is the index of a cell $B_\alpha$ used to define~$B$, and in
all the other cells we compute $\pi^A_\alpha$ via
\begin{equation}
  \label{eq:peqA}
  \sum_{\substack{\beta=1\\ \beta \not=\alpha}
  }^{\Lambda} \pi^A_{\beta} \nu^A_{\beta,\alpha} = 
  \sum_{\substack{\beta=1\\ \beta \not=\alpha}}^{\Lambda} 
  \pi^A_{\alpha} \nu^A_{\alpha,\beta}
\end{equation}
where the index $\alpha$ runs over all the cells outside of $A$ and
$B$ and we use as boundary conditions the values for
$\nu_{\beta,\alpha}^A$ and $\pi^A_\beta$ set before when the index
$\beta$ is that of a cell used to define~$A$ or $B$.
Finally, we compute the probability of re-entry on the edges of
a cell $B_\alpha$ not inside $A$ and $B$ as
\begin{equation}
  \label{eq:probfluxA}
  \PP^A_{\partial B_{\beta} \cap \partial B_{\alpha}}=
  \frac{\pi^A_{\beta} \nu^A_{\beta,\alpha}}
  {\sum_{\beta'\neq\alpha} \pi^A_{\beta'} \nu^A_{\beta',\alpha} }, 
  \quad (\beta\neq \alpha).
\end{equation}
This procedure automatically guarantees that we focus on the
trajectories assigned to $A$ and makes all the quantities above -- the
set of trajectories, the databanks, $\nu^A_{\alpha,\beta}$ and
$\pi^A_\alpha$ -- different from their unbiased statistical steady
state counterparts. By construction we then have $\pi^A_\alpha =
\int_{B_\alpha} \varrho_A(\zz) d\zz$ which, from~\eqref{eq:rhoAB},
means that
\begin{equation}
  \label{eq:piA}
  \rho_A = \sum_{\alpha=1}^\Lambda \pi_\alpha^A\le 1, \qquad \rho_B = 1 -\rho_A.
\end{equation}
Similarly we can compute $\nu_R$ from the total flux into $B$:
\begin{equation}
  \label{eq:nuRflux}
  \nu_R = \sum_{\substack{\text{$\alpha$ such that}\\ 
      \text{$B_\alpha$ not in $B$}}} \ \ 
  \sum_{\substack{\text{$\beta$ such that}\\ 
      \text{$B_\beta$ in $B$}}}
  \pi^A_\alpha \nu_{\alpha,\beta}^A.
\end{equation}
Finally, we can get $k_{A,B}$ and $k_{B,A}$ from~\eqref{sec:nuRkABkBA}
and be done.

\paragraph*{Comparison with TIS, FFS, and Markovian milestoning with
  Voronoi tessellations.} Before illustrating our procedure to compute
the reaction rate on a simple example, let us compare it to existing
methods. First, let us point out that it is very similar in spirit to
both TIS~\cite{tis2,pptis,tis3} and FFS\cite{ffs,valerianiFFS}. Like
TIS and FFS, our method is based on selecting only those trajectories
which come from the reactant state~$A$. The difference is in the way
this selection is achieved. In particular, unlike in TIS and FFS, we
do not require that the interfaces be ordered monotonously, i.e.~we do
not need that a trajectory coming from~$A$ crosses all the preceding
interfaces before reaching the next one. This offers more flexibility
in the way the interfaces can be chosen. Here we did so using the
edges of cells in a Voronoi tessellation because it is convenient, but
the formalism above is clearly independent of that choice and can be
applied to any type of interfaces.

Regarding the relation with the Markovian milestoning procedure with
Voronoi tessellation~\cite{miles6}, the main advantage of the method
proposed in this note to compute the rate is that it is exact and
hence avoids completely the assumptions underlying
milestoning~\cite{miles5}. On the other hand, the new procedure is
also more costly since it requires not only to build databanks of
re-entry points but also to do the sampling twice -- once to get the
unbiased statistical steady states quantities, and once more to get
the reaction rate by twisting the dynamics. For systems at
equilibrium, one may therefore prefer to use the Markovian milestoning
procedure with Voronoi tessellation~\cite{miles6} which is cheaper.
For nonequilibrium systems, this procedure is inapplicable (since it
relies on the time-reversibility of the dynamics), but as a
compromise, one could use the nonequilibrium sampling strategy for the
unbiased system to compute the relevant quantities in Markovian
milestoning and thereby avoid to make the second twisted sampling to
compute the rate. Indeed, the formalism developed in
Ref.~\onlinecite{miles6} to approximate the dynamics by a
continuous-time Markov chain does not rely on the dynamics being at
equilibrium. For completeness let us briefly recall the main objects
in Markovian milestoning and indicate how to compute them in the
present context. If, following Ref.~\onlinecite{miles6}, we define the
milestones as the common boundaries between any two adjacent Voronoi
cells and denote these milestones by $S_i$ with $i=1,2,\dots,N$, the
key quantity to approximate the transitions between the milestones by
a continuous-time Markov chain is the rate matrix whose off-diagonal
elements can be estimated as
\begin{equation}
  \label{eq:qMLE}
  q_{ij} = 
  \begin{cases}
  N_{ij}/R_i \quad &  \text{if $R_i\ne 0$}\\
 \quad 0  \quad &\text{if $R_i = 0$}.
  \end{cases}
\end{equation}
where 
\begin{equation}
  \label{eq:NijRi}
    N_{ij} = 
    \sum_{\alpha=1}^{\Lambda}\pi_{\alpha} \frac{N_{ij}^{\alpha}}{T_\alpha}, \qquad
    R_{i} = 
    \sum_{\alpha=1}^{\Lambda} \pi_{\alpha} \frac{R_{i}^{\alpha}}{T_\alpha}. 
\end{equation}
Here $T_{\alpha}$ is the total simulation time in cell $B_{\alpha}$
(as before), whereas $N_{ij}^\alpha $ is the total number of times the
trajectory went from milestone $S_i$ to milestone $S_j$ and
$R_i^\alpha$ is the total amount of time the trajectory is assigned to
milestone $S_i$ in cell $B_\alpha$, i.e.~the total amount of time this
trajectory is such that $S_i$ was the edge of cell $B_\alpha$ it hit
last. We refer the reader to Ref.~\onlinecite{miles6} for more
details.

\begin{figure}[t]
   \centerline{\includegraphics[width=3.25in]{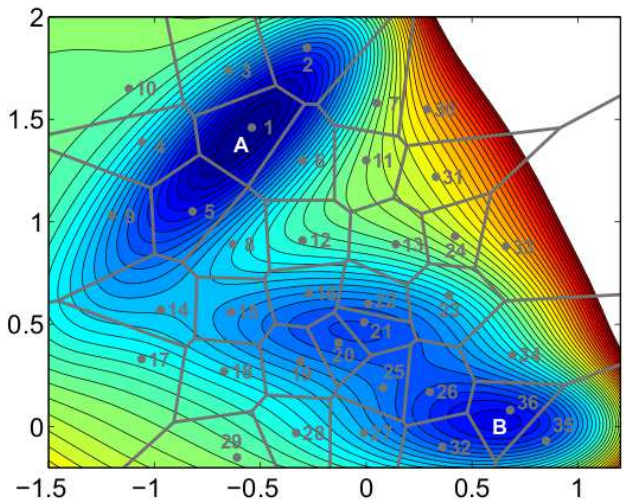}}
   \caption{%
     Contourplot of the Mueller potential with the 36
     points (shown as grey dots) generating the
     Voronoi tessellation shown as grey lines.
     The reactant $A$ and product $B$ states are
     identified with cell $B_1$ and $B_{36}$, respectively.
   }
\label{fig:potMull}
\end{figure}

\paragraph*{Illustrative example.} We end this note by illustrating
our sampling procedure on the example of a system evolving by Langevin
dynamics on a two-dimensional potential,
i.e.~$\zz(t)=(\xx(t),\vv(t))$, $\xx,\vv \in \RR^2$ and
\begin{equation}
  \label{eq:langevin}
  \begin{cases}
    \dot \xx(t) = \vv(t), \\
    \dot \vv(t) = - \nabla V(\xx(t)) - \gamma \vv(t) +
    \sqrt{2\beta^{-1}\gamma}\, \eeta(t).
  \end{cases}
\end{equation}
Here $V(\xx)$ is the Mueller potential~\cite{mueller} whose
contourplot is shown in Fig.~\ref{fig:potMull}, $\beta=1/(k_BT)$ is
the inverse temperature and $\eeta(t)$ is a Gaussian white-noise with
mean zero and covariance $\< \eta_i(t) \eta_j(t')\> = \delta_{ij}
\delta(t-t')$. $\gamma$ is the friction coefficient and, for
simplicity, we have set the mass tensor to the identity. Below we took
$\gamma=100$ and $\beta^{-1}=20$ (which is about $20\%$ the value of
the energy barrier between the minimum at the top left corner of the
Mueller potential and the saddle point at (-0.8,0.6)). This example is
obviously very simple and serves no other purpose than being a
benchmark: the application of our sampling procedure to more
interesting examples will be reported elsewhere. To avoid confusions,
before presenting our results for~(\ref{eq:langevin}) let us note that
this system is an equilibrium one, with equilibrium density
$\varrho(\xx,\vv)= Z^{-1} \exp(-\beta H(\xx,\vv))$ where $H(\xx,\vv) =
\frac12 |\vv|^2 + V(\xx)$ is the Hamiltonian and $Z =
\int_{\RR^2\times\RR^2}\exp(-\beta H(\xx,\vv))d\xx d\vv$ the partition
function. We are, however, primarily interested in computing the
reaction rates between the reactant state $A$ and the product state
$B$ shown in Fig.~\ref{fig:potMull}, which we will achieve by twisting
the dynamics as explained before to focus on trajectories assigned to
$A$. In so doing, we automatically put the system out of equilibrium
since $A$ becomes a source and $B$ a sink (and $\varrho_A(\xx,\vv)
\not = \varrho(\xx,\vv)$). This is why we need the nonequilibrium
formalism developed above to compute reaction rates even in the case
of an equilibrium system such as~(\ref{eq:langevin}).

Shown in Fig.~\ref{fig:potMull} are the 36 cells that we used in our
calculations. These cells were defined as
\begin{equation}
  \label{eq:Bkx}
  \begin{aligned}
    B_{\alpha}=
    \{(\xx,\vv) \in \RR^2\times\RR^2 \; : \;
    |\xx - &\xx_{\alpha}|<|\xx-\xx_{\beta}| \\
    & \text{for all} \; \beta \ne \alpha \},
  \end{aligned}
\end{equation}
where $|\cdot|$ denotes the Euclidean norm and $\xx_\alpha$ with
$\alpha=1, \ldots, 36$ are points chosen randomly in the region where
the energy is below a certain threshold value. We identified the
reactant and product states with two single cells in the neighborhood
of the two deep minima in the potential landscape: $A=B_1$ and
$B=B_{36}$ (see Fig~\ref{fig:potMull}). Note that we intentionally
took many cells and disposed them in a way that may not be optimal for
the reaction to check that our procedure is robust against such
choices. Note also that by defining the cells primarily in position
space as we do in~(\ref{eq:Bkx}) we guarantee that we can count
successive crossing of their boundaries. This would not be the case if
the boundaries of the cells were bent in velocity space, because of
the noise term in~(\ref{eq:langevin}) that acts on the velocities.

Both steps of the sampling procedure (the one to compute the unbiased
equilibrium quantities and the other with the twisted dynamics) were
performed as described above by running simulations for $10^8$ steps
in each cell using the second order integrator of
Ref.~\onlinecite{langevin} with a timestep $\mathit{\Delta
  t}=10^{-4}$. In the first step of the procedure we took $\pi_\alpha$
equal in each cell as initial condition; in the second step, we took
$\pi_\alpha^A = \pi_\alpha$ as initial condition. We compared the
results of our procedure with those obtained by generating a long
($10^{10}$ steps) unbiased trajectory by brute force simulation and
computing $\nu_R$, $k_{A,B}$ and $k_{B,A}$ from finite~$T$
approximations of the limits in~(\ref{eq:nuAB}) and~(\ref{eq:kAB}).

\begin{table}[th]
  \centering
  \begin{tabular}{|l|c|c|c|}
    \hline
                                    &  $\nu_R$               &   $k_{A,B}$    &  $k_{B,A}$     \\
    \hline
    Twisted dynamics   &  $5.9 \,10^{-3}$   &   $7.4 \,10^{-3}$  &  $3.1\,10^{-3}$          \\
    Direct simulation    &  $5.8 \,10^{-3}$   &   $7.1 \,10^{-3}$  &  $3.2\,10^{-3}$          \\
  \hline
  \end{tabular}
  \caption{%
    Rate of reactive trajectories and reaction rates between the
    sets $A$ and $B$ in the Mueller potential (see Fig.~\ref{fig:potMull})
    obtained by the procedure presented here and by direct calculation
    using a long unbiased trajectory. 
 }
 \label{tab:rates}
\end{table}
\begin{figure}[t]
  \centerline{\includegraphics[width=3.25in]{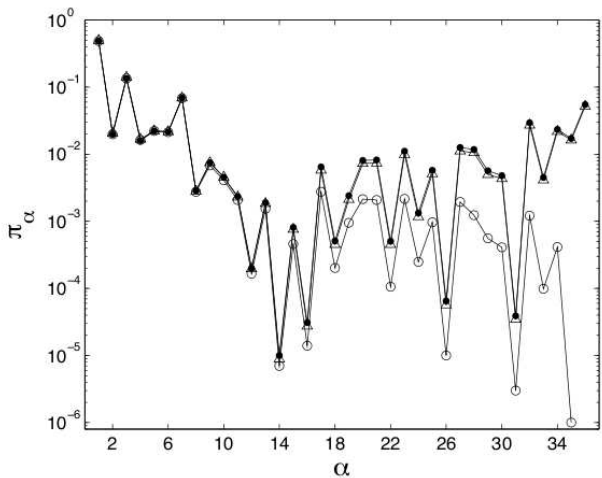}}
\caption{%
 Probability $\pi_\alpha$ to be in the cells (semi-log scale)
 obtained via sampling restricted in the Voronoi cells (black dots)
 and by a brute force unbiased trajectory (triangles). 
 The agreement is excellent. 
 Also shown as circles are the steady state nonequilibrium 
 probabilities $\pi^A_\alpha$ of the twisted dynamics. 
 Note that by construction $\pi^A_{36}=0$ 
 (since $B_{36}=B$) and we did not plot this point. }
\label{fig:comparisonpi}
\end{figure}

The main outputs of our procedure are the rate of reactive
trajectories $\nu_R$ and the reaction rates $k_{A,B}$ and $k_{B,A}$
which are reported in Table~\ref{tab:rates}. These are within
statistical errors of the corresponding quantities estimated by brute
force simulation. Our procedure also produces the equilibrium
probabilities $\pi_\alpha$: as shown in Fig.~\ref{fig:comparisonpi}
these are within statistical errors of the corresponding values
obtained by the brute force simulation. Also shown in
Fig.~\ref{fig:comparisonpi} are the probabilities $\pi_\alpha^A$
computed by twisting the dynamics. As expected, the closer to $B$, the
smaller $\pi_\alpha^A$ is compared to $\pi_\alpha$.

\paragraph*{Acknowledgements.} This work was motivated by a question
from Benoit Roux. Partial support by NSF grants DMS02-09959,
DMS02-39625 and DMS07-08140, and ONR grant N00014-04-1-0565 is also
acknowledged.

\end{document}